\documentclass[%
reprint,
superscriptaddress,
nofootinbib,
amsmath,amssymb,
prd,
]{revtex4-2}

\usepackage{amsmath}

\usepackage{amssymb}
\usepackage{amsfonts}
\usepackage{mathrsfs}
\usepackage{mathtools}

\usepackage[utf8x]{inputenc}
\usepackage{slashed}
\usepackage{xcolor}
\usepackage{caption}
\usepackage{subcaption}
\usepackage[T1]{fontenc}
\usepackage[english]{babel}
\usepackage{amsmath}
\usepackage[english]{babel}
\usepackage{nicefrac}
\usepackage{braket}

\usepackage{graphicx}
\usepackage{float}
\usepackage{hyperref}
\definecolor{unamblue}{cmyk}{1 0.79 0.12 0.59}

\usepackage[]{hyperref}
\hypersetup{
	colorlinks=true,%
	citecolor=unamblue,%
	filecolor=unamblue,%
	linkcolor=unamblue,%
	urlcolor=unamblue,
	bookmarksnumbered=true,     
	bookmarksopen=true,         
	bookmarksopenlevel=1,       
	pdfstartview=Fit,           
	pdfpagemode=UseOutlines,
	pdfpagelayout=TwoPageRight
}
\usepackage{comment}

\usepackage[]{todonotes}

\usepackage{graphicx}
\usepackage{tikz}
\usepackage[com pat=1.1.0]{tikz-feynman}
\usepackage{circuitikz}
\usetikzlibrary{decorations.pathmorphing}
\usetikzlibrary{shapes.geometric}
\usetikzlibrary{matrix, calc, arrows}
\tikzset{snake it/.style={decorate, decoration=snake}}
\tikzset{gluon/.style={decorate,
		decoration={coil,amplitude=3pt, segment length=4pt,  pre length=.1cm, post length=.01cm}}
}

\usetikzlibrary{decorations.markings}

\tikzset{%
	dots/.style args={#1per #2}{%
		line cap=round,
		dash pattern=on 0 off #2/#1
	}
}

\tikzfeynmanset{
	sdot/.style={
		/tikzfeynman/dot,
		/tikz/minimum size=3pt,
	},
	gluon/.style={decorate,
		decoration={coil,amplitude=3pt, segment length=4pt,  pre length=.1cm, post length=.01cm}
	},
	graviton/.style={decorate, double,
		decoration={snake}
	},
	compact arrow/.style={
		/tikzfeynman/momentum/.cd, arrow distance=8pt, label distance=-6pt, arrow shorten=0.27
	},
}

\newcommand{\be }{\begin{equation}}
	\newcommand{\ee }{\end{equation}}
\newcommand{\ba }{\begin{equation} \begin{aligned}}
		\newcommand{\ea }{ \end{aligned} \end{equation} }

\newcommand{\deltahat}{\hat{\delta}}

\newcommand{\dd}{\mathrm{d}}
\newcommand{\ii}{\mathrm{i}}
\newcommand{\hKS}{{\tilde {h}^{\textrm{KS}}}}

\newcommand{\xo}{\mathbf{x}_0}

\begin{document}
	
	\title{Gravitational lensing in a  plasma from worldlines}
	
	\author{Francesco Comberiati}
\affiliation{Dipartimento di Scienze Economiche,
 Universit\`a di Bologna, 	Piazza Scaravilli 2, I-40126 Bologna, Italy}

\author{Leonardo de la Cruz}

	\affiliation{Institut de Physique Th\'eorique,  CEA, CNRS, Universit\'e Paris–Saclay, F-91191, Gif-sur-Yvette cedex, France}

	%
	\begin{abstract}
		We study  the deflection of light rays  in a cold, non-magnetized plasma using the  worldline framework. Starting from 
		Synge's Hamiltonian formalism, 
		we construct a position-space action and use it perturbatively to calculate light bending angles. In the homogeneous case, the action reduces to that of a massive particle, allowing us to extract the bending angle of light in the presence of the medium using a well-known analogy. For the inhomogeneous case, we consider a power law model and construct Feynman rules in time to compute the purely plasma-induced corrections to the bending angle at Next-to-Leading-Order (NLO). 
	\end{abstract}
	\maketitle
	\section{Introduction}
	In the presence of an optical medium the bending  of light rays
	predicted by Einstein's general relativity in vacuum is modified to account for the characteristic frequency of the medium and on that of light itself.	In the geometrical optics approximation, light in  an isotropic non-dispersive  medium  follows a geodesic equation with a modified metric that depends on the properties of the medium encoded in a refraction index \cite{Gordon:1923qva, Ehlers+1967+1328+1332}.
	 A self-consistent approach to 
	treat light propagation in the gravitational field in a medium
	  was developed by Synge in his classic textbook~\cite{Synge:1960ueh}. The influence of the medium in the bending of light is relevant for investigations of radio waves near the solar corona \cite{PhysRevLett.17.455}, gravitational lensing~\cite{Virbhadra:1999nm,Virbhadra:2008ws, Bisnovatyi-Kogan:2017kii}, and shadows cast by black holes~\cite{Cunha:2018acu,Perlick:2021aok}.

The deflection angle of light in a  plasma in Schwarzschild has been studied in detail by Bisnovatyi-Kogan and Tsupko \cite{Bisnovatyi-Kogan:2008qbk, Bisnovatyi-Kogan:2010flt} and by Perlick \cite{Perlick2000} for Kerr. In vacuum, the weak deflection angle
 can also be computed through the Gau{\ss}-Bonet theorem
 as shown by   Gibbons and Werner \cite{Gibbons:2008rj}. This remarkable idea was used by 	Crisnejo and  Gallo~\cite{Crisnejo:2018uyn} to compute the deflection angle in a plasma and later used to obtain higher order corrections~\cite{Crisnejo:2019ril}. The effects of plasma in gravitational lensing have been quantitatively studied in Refs.~\cite{Er:2013efa,Kimpson:2019mji,Kimpson:2019hwn,Er:2022lad}. 

In a different context, a  program aimed toward applying quantum field theory (QFT) amplitudes  to describe classical scattering has 
gained significant attention \cite{Bjerrum-Bohr:2022blt, Kosower:2022yvp}. 
 In particular, Feynman's idea of describing QFT amplitudes as relativistic \emph{worldline} path integrals is one of the techniques that allows a smooth transition from  quantum to classical physics as was recognized long  ago~ \cite{Fabbrichesi:1993kz}.  More recently, the application of worldline methods\footnote{The wordline approach has been reviewed in Refs.~\cite{Schubert:2001he,Bastianelli:2005rc, Edwards:2019eby, Bastianelli:2025, Porto:2016pyg,Levi:2018nxp}} to effectively compute
classical observables in the post-Minkowskian~(PM) expansion was   considered by Mogull, Plefka, and  Steinhoff in Ref.~\cite{Mogull:2020sak} and further developed in Refs.~\cite{Jakobsen:2021smu, Jakobsen:2021zvh,Shi:2021qsb,Jakobsen:2022psy,Bastianelli:2021nbs}. This approach is  formally equivalent to the so-called
  Worldline Effective Field Theory (WEFT) framework to gravitational dynamics~\cite{Goldberger:2004jt,Kalin:2020mvi, Liu:2021zxr, Kalin:2022hph}. In both cases, these can be related \cite{Comberiati:2022ldk,Damgaard:2023vnx,  Capatti:2024bid} to QFT observables,   defined from the $S$-matrix \cite{Kosower:2018adc,Maybee:2019jus,delaCruz:2020bbn}.
  
  Generally,   the goal of the worldline approach is to construct a generating function (partition function) of connected correlation functions 
	\begin{align}
	\mathcal Z = \int \mathcal{D}x \, 
	e^{\ii S[x;g]}\, ,
	\label{path-integral}
\end{align}   
where we integrate over all possible  trajectories $x^\mu(\tau)$. For scattering of two \cite{Fabbrichesi:1993kz} or more \cite{Feal:2022iyn, Feal:2022ufw} worldlines, one also integrates over the background  fields.
The classical limit is  then obtained as usual by taking expectation values of certain operators and restricting to tree-level diagrams  \cite{Das:2019jmz}.  While this framework is  unambiguous in flat space, constructing path integrals in curved spacetime unambiguously is  nontrivial due to the necessity of introducing ultraviolet regulators
 \cite{Bastianelli:2002fv, Bastianelli:2005vk, Bastianelli:2006rx}. 
  In the classical limit, these can be neglected and the formalism becomes simpler. In Ref.~\cite{Bastianelli:2021nbs}, the current authors and Bastianelli, considered  light bending in the worldline formalism by first constructing the full photon propagator dressed by gravitons and later restrict the analysis to the geometrical optics regime.

	In this paper, we will consider light bending in an optical medium using the worldline approach. This paper is organized as follows. In section~\ref{worldlinesinthemedium}, 	we construct an action in position space suitable to apply worldline methods. In the homogeneous case the problem reduces to the probe limit, which we analyze in detail in section~\ref{problelimit}.  In section~\ref{nonhomogeneous}, we  examine the inhomogeneous case and calculate the first-order correction to the deflection angle induced by the plasma. Our conclusions are presented in section~\ref{conclusions}.

		\newcommand{\VV}{V}
	\newcommand{\spx}[1]{\mathbf{#1}}
	\section{Worldlines in the medium}
	\label{worldlinesinthemedium}
	We will use the mostly minus signature for the Minkowski metric $\eta_{\mu\nu}=\text{diag}(1, -1,-1,-1)$
	and set the gravitational coupling to  $\kappa=\sqrt{32\pi G}$, where $G$ is the Newtonian constant.  The gravitational action is given by the usual Einstein-Hilbert action 
	\begin{align}
		S_{\text{EH}}[g]=- \frac{2}{ \kappa^2} \int \dd^4 x \sqrt{-g}R \;. 
	\end{align}
Our starting point is Synge's framework for  geometrical optics in a medium~\cite{Synge:1960ueh}.  By analogy with classical optics Synge introduces a refraction index $n(x)$ 
	of the medium
	\begin{equation}
		n^2(x)= 1-\frac{p_\alpha p^\alpha}{
			\omega(x)^2} \, ,
	\end{equation}
	which depends
	on the photon frequency $\omega(x)=p_\mu \VV^\mu$  in the medium,  and the space-time coordinates $x$. Here  $\VV^\mu$ is the 4-velocity of the medium that plays the role of the observer.  The formalism assumes that the refraction index is known. The Hamiltonian  of a general dispersive medium is given by 
	\begin{equation}
		H(x,p)= \frac 12  \bar{g}^{\mu\nu} p_\mu p_\nu \, ,
	\end{equation}
	where the  tensor $\bar g^{\mu\nu}$ reads
	\begin{equation}
		\bar{g}^{\mu\nu}
		= g^{\mu\nu}+(n^2-1)\VV^\mu \VV^\nu\, .
	\end{equation}
Its inverse 
	\begin{equation}
		\bar g_{\mu\nu}= g_{\mu\nu}-\left(1-\frac {1}
		{n^2} \right)
		\VV_\mu\VV_\nu \, 
	\end{equation} 
	 satisfies  $\bar g 	_{\alpha\mu} \bar g^{\alpha\nu}
	=\delta_\mu^{\ \nu}$. It is only for non-dispersive media (when the refraction index $n$ is a function of the position only) that the tensor $\bar{g}^{\mu\nu}$ can be regarded as a metric tensor~\footnote{See Ref.\cite{Crisnejo:2019xtp} for a recent discussion about this point.}. We will assume this case below.	 Under the assumptions of a static space-time the vector  $\VV^\mu$ can be taken, say, as
	$\VV^\mu=g_{00}^{-1/2}(1,0,0,0)$. Within the spacetime, we consider a cold non-magnetized plasma where the  refraction index is
	\begin{equation}
		n_e^2= 1-\frac{\omega_e(x)^2}{
			\omega(x)^2} \, ,
	\end{equation}
	 where $\omega(x)$ is the  photon frequency 
	and $\omega_e(x)$ is  the electron plasma frequency. The latter is related to the electron density function $ N(x)$ by
	\begin{equation}
		\omega_e(x) = \sqrt{\frac{4\pi e^2}{m_e} N(x)} \, , 
	\end{equation}
	where $e$,  $m_e$, are the charge, and the mass of the electron, respectively.
	 Then, the Hamiltonian can be written as
	\begin{equation}
		\begin{aligned}
			H=
			\frac{1}{2} \left(g^{\mu\nu}-
			\frac{\omega_e^2}{(p_\mu \VV	^\mu)^2} \VV^\mu \VV^\nu \right)
			p_\mu p_\nu
			\\= \frac{1}{2} \left(
			g^{\mu\nu}p_\mu p_\nu-
			\omega_e^2
			\right)\, .
		\end{aligned}
	\end{equation}
	We may thus interpret this Hamiltonian as that of a massless particle and a potential $\omega^2_e(x)$. The  particle action in phase space then reads
	\begin{equation}
		S[x,p; g]= -\frac{1}{2}
		\int\dd \tau  \,  \left[
		p_\mu \dot x^\mu-
		\frac {e(\tau)} {2}\left(g^{\mu\nu}p_\mu p_\nu -\omega_e^2\right)
		\right] \, ,
		\label{action-position-space.general}
	\end{equation}  
	where we have introduced 
	the einbein $e(\tau)$, which accounts for invariance under reparametrizations of the worldline. 
	We will gauge fix it as $e(\tau)=1/\omega(-\infty)\equiv 1/\omega_0$ to be the asymptotic frequency of the photon. Unless stated otherwise, the integration limits will be over the real line. The action in position-space then reads 
	\begin{equation}
		S[x;g]=
		-\frac{1}{2} \omega_0
		\int\dd \tau  \,  \left[
		g_{\mu\nu} \dot x^\mu\dot x^\nu+ \frac{\omega_e(x)^2}{\omega_0^2}
		\right] \, ,
		\label{action-position-space}
	\end{equation}
	which, in principle, can be used to quantize the theory using path integrals (see Ref.~\cite{Bastianelli:2005rc} for an overview). A similar action has previously been studied in the context of causality studies \cite{Hollowood:2007ku,Hollowood:2007kt}.
	
However, our interest  lies in perturbative classical observables within this formalism \cite{Mogull:2020sak}. In this setup, one considers a background expansion of the trajectory $x^\mu$ together  with a weak field expansion of the metric around flat space ($g_{\mu\nu}=\eta_{\mu\nu}+\kappa h_{\mu\nu}$). The background expansion of the trajectory reads
	\begin{equation}
		x^\mu(\tau)= x_0^\mu(\tau) + z^\mu(\tau)   \, , \label{splitting-straight-line}
	\end{equation}
	where $x_0^\mu=b^\mu+u^\mu \tau$ is the straight line trajectory, and  $u^\mu$ is the velocity of the photon in the medium at $\tau=-\infty$.  We also assume that $\omega_e(\infty)=\omega_{e0}=\text{const}$. 
	This means that even in flat space the momentum squared 	of the photon is different from zero. In flat space, we have  $ p^\mu= \omega_0(1, n_{e0} \hat{\mathbf{e}})$, with $n_{e0}^2=1-\omega_{e0}^2/\omega_0^2$ and $\hat{\mathbf{e}}$ a unit spatial vector. Photons in a plasma can then be characterized by an effective mass $m_{\text{eff}}=\omega_0 \sqrt{1-n_{e0}^2}=\omega_{e0}$. For our purposes it would be useful to parametrize the photon momentum as
	$p^\mu = m_{\text{eff}} u^\mu$ with
	\begin{equation}
		u^\mu= \frac{\omega_0}{m_{\text{eff}}}(1, n_{e0} \hat{\mathbf{e}})    \, .
	\end{equation}
We then have  $u^2=1$.  In contrast,  the effective velocity in a plasma reads
	\begin{equation}    u^2_{\text{eff}}=1 -\frac{\omega^2_{e0}}{\omega^2}
		=n_{e0}^2 \, ,
	\end{equation}
	which corresponds to its group velocity. 
	One can also write the above quantities for any stationary gravitational field 
	\cite{Bisnovatyi-Kogan:2010flt} but we will not need them here.
	
	In the homogeneous case, the electron density $N(x)$ is constant so the second term in Eq.~\eqref{action-position-space} is  also a constant.   We can think of it as an effective  mass
	as we did above in flat space. This well-known 
	analogy  can be used to map scattering angles computed in the scenario of a photon in a homogeneous plasma and the probe limit of a massive particle in vacuum \cite{Bisnovatyi-Kogan:2015dxa}. Therefore, in principle  we could apply the worldline approach of Ref.~\cite{Mogull:2020sak} to compute observables for scattering and then take probe limit defined as taking the mass of one of the particles in the scattering to be very large. In this limit,   the velocity of the probe at infinity enters through the  Lorentz factor $\gamma=1/\sqrt{1-v^2}$. The medium then enters through the relation
	\begin{equation}
		v \to u_{\text{eff}}=
		\sqrt{1-\omega_{e0}^2/\omega_0^2} \, .
		\label{maptomedium}
	\end{equation}
	 In fact, 
	we could use any other available approach  to describe the probe limit(e.g. \cite{Kol:2021jjc,Menezes:2022tcs, Damgaard:2022jem, Hoogeveen:2023bqa}), and with  this map obtaining results in the homogeneous case. However, we will instead develop a direct worldline-based approach that inherently implements this limit and applies to the homogeneous case. Let us now consider the probe limit.
	\section{Probe limit and worldlines}
	\label{problelimit}
	In 	the probe limit,  we neglect the black hole recoil due to the emission of gravitons, thus keeping it as a fixed object. This means that the background metric is maintained as some fixed solution to the Einstein field equations, representing a black hole.  We then analyze a worldline for the probe as it scatters off this background.
	
	Our goal in this section is to provide different techniques to solve the worldline path integral in this scenario, in order to extract the probe deflection angle in the background.
	We consider one massive (the massless probe can be treated in a similar manner) scalar particle scattering off a black hole background. In calculations, 
		we  
	move to the reference frame of the black hole parametrizing its momentum as $V^\mu = M (1,0,0,0)$, where $M$ is the black hole mass.
	The massive probe is aligned along the $z-$axis  
	so its momentum is $p^\mu = m\gamma (1,0,0,v)$.
	
	\newcommand{\vv}{\beta}
		The worldline action describing the motion of a scalar particle in a generic gravitational background is 
	\be \label{worldline-action}
	S[x;g] = -\frac{1}{2}\int\dd\tau \, g_{\mu\nu}(x(\tau)) \dot{x}^\mu(\tau)\dot{x}^\nu(\tau) \,, 
	\ee
	while the related one particle partition function can be written simply as Eq.~\eqref{path-integral}.
	This path integral  is translational invariant
	namely  $\mathcal{D}x=\mathcal{D}(x_{0}+z)=\mathcal{D}z$.  
	Here  $x_{0}$ is not restricted to be the straight line but can be any classical solution of the equations of motion. 
	In comparison to the action Eq.~\eqref{action-position-space},  Eq.~\eqref{worldline-action} can be obtained by rescaling\footnote{See Ref.~\cite{Bastianelli:2021nbs} for the analogous case of the photon.} it as $\tau\to \tau /\omega_0$  and the constant mass term integrated out in  the definition of $ \mathcal Z$. Thus, in our background expansion we will have $\dot x^\mu=p^\mu$.
The observables  are 
 defined by
	\begin{equation}
		\braket{\mathcal O}:= \frac{1}{\mathcal Z} \int \mathcal Dx \ \mathcal{O} e^{\ii S[x;g]}  \, ,
	\end{equation}
	which we can compute using Wick contractions. 	The perturbation theory is controlled by the fluctuation $z^\mu(\tau)$. The basic Wick contraction of two fields defines  the retarded propagator
 \begin{equation}
 	\braket{z^\mu(\tau_1) z^\nu(\tau_2)}= G^{\mu\nu}(\tau_1-\tau_2), 
 \end{equation}
 whose explicit form is
	\begin{equation}
	 G^{\mu\nu}(\tau_1-\tau_2)= \frac{-\ii \eta^{\mu\nu}}{\mathcal {E}}  \left(|(t_1-t_2)|+(t_1-t_2)\right),
	 \label{time-domain-propagator}
	\end{equation}
	where $\mathcal E$ is the overall factor that appears in the action ($\omega_0$ in the general case and unity in this section). It is often convenient to work in Fourier space, where
	\begin{equation}
		\begin{aligned}
		z^\mu(\tau)=\int\hat{\dd}\omega \  e^{\ii \omega \tau}
		\tilde z^{\mu}(-\omega), \\
			f(x)=\int\hat{\dd}^4 q\  e^{\ii q\cdot x}
	\tilde f(-q),
			\end{aligned}
		\label{convention-fourier}
	\end{equation}
	where we have introduced the short-hard notation $\hat{\dd}^n x=\dd^n x/(2\pi)^n$ to absorb factors of $2\pi$. We also define $\hat \delta (x)=2\pi \delta(x)$. In energy space, the  propagator reads
	\begin{equation}
	\tilde	G^{\mu\nu}(\omega_1,\omega_2)
		=-\ii \frac{\eta^{\mu\nu}}{\mathcal{E}} \frac{ 
			\delta(\omega_1+\omega_2)}{
			\omega_1^2}.
			\label{propagator-energy-space}
	\end{equation}
	Let us now focus on the background expansion of $x^\mu(\tau)$. The following discussion is standard yet clarifying in our context~\cite{Das:2019jmz}. We split the configuration space variables as 
	\begin{equation}
		x^\mu(\tau)=
		x^\mu_0(\tau)+z^\mu(\tau) \,,
		\label{usual-split}
	\end{equation}
	where $x_0^\mu$ is a solution of the classical equations of motion (e.o.m.) of the point particle  originating from \eqref{worldline-action}.
Let us  perform a Taylor expansion of the action in powers of $z^\mu$ around the  classical solution $x_0^\mu$
\be \label{expansion-action}
S[x_0+z;g] = \sum_{n=0}^{\infty} \frac{1}{n!}
{S_{\mu_1 \dots \mu_n}}[x_0]z^{\mu_1}z^{\mu_2}\cdots z^{\mu_n} \,,
\ee
where the expansion coefficients are defined by
\begin{equation}
	{S_{\mu_1 \dots \mu_n}}[x_0]
	:=\frac{\delta^n S[x_0+z]}{\delta z^{\mu_1}\delta z^{\mu_2} \cdots \delta z^{\mu_n}} \Big|_{z=0} \, . 
	\label{coefficients}
\end{equation}	
\subsection{Schwarzschild}
\label{Schwarzschild}
Let us consider Schwarzshild and 
 use  the saddle point approximation of the path integral to evaluate the partition function. This yields
\be 
{\cal Z} = \textrm{Det} \left(
\frac{\delta^2 S}{\delta z^2} \right)\,\exp\left( \ii S[x_0]\right)\,.
\label{saddle-point-approximation}
\ee
Notice that having kept the background to be fixed and expanding around an exact solution $x_0$, the above expression is  the exact value of the partition function. The connected components of the partition function are $\mathcal W=-\ii \log\mathcal Z$, which we can identify with the action evaluated at the extreme points. For the WKB analogous see Ref.~\cite{Kol:2021jjc}.

 In order to evaluate explicitly the action on the exact trajectory, we use the Hamilton-Jacobi method. The Hamilton principle function in a Schwarzshild space-time is $p_\mu =\dot{x}_\mu =  \partial_\mu W $. Using spherical coordinates, it can be separated into the components
 \be 
 W(t,r,\theta,\varphi) = E t+ W_r(r) + W_\theta(\theta) + L_\varphi\, \varphi \,. 
 \ee
 In the equatorial plane $\theta = \frac{\pi}{2}$, the evaluation of the path integral leads to 
 \ba
 {\cal Z} = & e^{-\frac{\ii}{2} \left(W_r(r=\infty)
 	-W_r(r_0)
 	\right)
 } \\
 & \times e^{  -\frac{\ii}{2}L_\varphi (\varphi(\infty) - \varphi_0)  
 } e^{-
 	\frac{\ii}{2}E \left(t(\infty)-t(-\infty) \right)
 } \, , 
 \ea
 where we absorbed the functional determinant in the definition of the partition function.
 As expected the partition function factorizes into the radial, angular, and time components. 
 The radial component of the Hamilton principle function for a massive particle moving in a Schwarzschild background is well known. It reads  
 \begin{align} 
 	W_r(r) =& \int_{r_0}^r \, dR\, \sqrt{U(R)}, \\
 	U(R) =& \frac{m^2 (\gamma^2 -1)R^2+2G m^2 M R- L^2 \frac{R-2GM}{R}}{(R-2GM)^2} \ , 
 \end{align}
 with $r_0$ being the largest zero of $U(R)$ and $L= b m v \gamma$. In slight abuse of notation we will use $b\equiv \sqrt{-b^\mu b_\mu}$.
 Using that $\mathcal W=-\ii \log\mathcal Z$, we can identify $ \mathcal W$ with the radial action $\chi_r$, thus finding
 \be 
 W_r(r=\infty)
 -W_r(r_0) = -2 \chi_r \, . 
 \ee
We are interested in  the map 
\eqref{maptomedium} so we will write our quantities as functions of $v$.
 The deflection angle is  derived as usual from
 \begin{widetext}
 	\ba
 	\alpha  &= \frac{d(-2 \chi_r)}{d L} = \int_{r_0}^\infty \dd r\, \frac{2 L}{r(r-2GM) \sqrt{U(r)}}\\
 	&= \pi +\frac{2 G M}{b v^2}\left(1+v^2\right)
 	+ \frac{3 \pi\ G^2 M^2}{4 b^2 v^2}\left(4+v^2\right)
 	+\frac{2 G^3 M^3 \left(5 v^6+45 v^4+15 v^2-1\right)}{3 b^3 v^6}\\
 	&+\frac{105 \pi G^4 M^4 \left(v^4+16 v^2+16\right)}{64 b^4 v^4}
 	+\frac{2 G^5 M^5 \left(21 v^{10}+525 v^8+1050 v^6+210 v^4-15 v^2+1\right)}{5 b^5
 		v^{10}}\\
 	&+\frac{1155 \pi G^6 M^6 \left(v^6+36 v^4+120 v^2+64\right)}{256 b^6 v^6}
 	+\dots  \, ,
 	\label{angle-order}
 	\ea
 \end{widetext}
 where we have integrated order by order in $G$.
 The integral has closed expressions in terms of the Appel $F_1$ function \cite{Kol:2021jjc} and in terms of elliptic functions \cite{Scharf:2011ii}. While a similar calculation can be performed for the Kerr case using the methods outlined in Ref.~\cite{Damgaard:2022jem}, we will instead explore an alternative approach.
 
 Before proceeding, however, it is worth discussing an important aspect of our approach: the application of the background field method for worldlines. 
 In the above calculations we used the background field on the worldline variables, keeping the gravitational field to be fixed. This allowed us to recover results in the probe limit for the scattering, as expected.
 An interesting scenario  would be to allow for fluctuations around the background metric. In such a case, the saddle point approximation would not be enough to recover results, since $x_0^\mu$ would not be the exact trajectory this time. However, one could set up the perturbative expansion of the path integral, expanding around the background metric and an exact solution in the background metric, for the graviton and worldline path integral respectively. This would lead to the so called self-force expansion, which would be interesting to study further. In the WEFT context this idea  was considered already  in Ref.~\cite{Cheung:2023lnj}.

	\subsection{The Kerr-Schild gauge}
In the previous subsection, we have used an exact solution of the e.o.m. to evaluate the deflection angle but one can also consider a weak-field expansion of the metric. In the case of Kerr, the idea is 
	to set up perturbation theory by expanding the metric tensor as
	\begin{equation}
		g_{\mu\nu}(x) = \eta_{\mu\nu} +  h^{\text{KS}}_{\mu\nu}(x)\, , 
		\label{splitting-1}
	\end{equation}
	where the  linearized Kerr background is chosen in Kerr-Schild (KS) gauge. Let us  introduce the spin of the black hole 
	\begin{equation}
		a_{\mu} = \frac{1}{2M}\varepsilon_{\mu\nu\rho\sigma}V^\nu S^{\rho\sigma} \, ,
	\end{equation}
	which obeys the supplementary spin condition (SSC), namely $a\cdot V = 0$.
	Let $J_n(x)$ be the Bessel functions of the first kind and $j_n(x)=\sqrt{\pi/(2x)}J_{n+1/2}(x)$. 
	The  Fourier-space linearized Kerr metric in KS gauge was computed in  Ref.~\cite{Bianchi:2023lrg}. Its covariant form can be written as
	\newcommand{\xx}{X}
	\begin{widetext}
		\begin{equation}
			\begin{aligned}
				\tilde{h}^{\text{KS}\mu\nu}(q)=
				8\pi G M &\Bigg[-\frac{j_1(\xx) \varepsilon^{\mu}(a,q,V) \varepsilon^{\nu}(a,q,V)}{ \xx | q| ^2}+\frac{2 \ii j_0 (\xx)  V^{(\mu }\varepsilon^{\nu)}(a,q,V)  
				}{M  |
					q| ^2}+\frac{\pi  J_1(\xx) q^{(\mu } \varepsilon^{\nu) }(a,q,V)}{\xx | q| ^3}\\
				&+\frac{  j_0(\xx)\left(-\frac{2 q^{\mu } q^{\nu }}{| q| ^2}-\eta^{\mu
						\nu }\right)}{| q| ^2}+\frac{V^{\mu }V^{\nu } \left(\frac{\sin (\xx)}{\xx}+\cos (\xx)\right)}{M^2 | q| ^2}-\frac{\ii \pi  J_0(\xx)
					q^{(\mu } V^{\nu) }}{M | q| ^3}\Bigg]\delta \left(q\cdot V\right)\, ,
			\end{aligned}
		\end{equation}
	\end{widetext}
	where $|x|:=\sqrt{-x^\mu x_\mu}$.
	We have also defined
		\begin{align}
			\varepsilon^{\mu} (a,q, V):=& \frac{1}{M}\varepsilon^{\mu \alpha\beta \nu}a_{\alpha}q_{\beta}V_{\nu}, \\ 
			X :=&  \sqrt{ a^2 q^2 - (a\cdot q)^2} \, .
		\end{align} 
	It is easy to verify that $\eta^{\mu\nu} \tilde h^{\text{KS}}_{\mu\nu}(q) = 0 $, as a consequence of the kinematics.  Notice that $\tilde {h}^{ \text{KS}\mu\nu} (q)$ is of order $\mathcal{O}(\kappa^2)$.
	In addition to the above expansion, we  split the configuration space variables as in Eq.\eqref{splitting-straight-line} with $x_0^\mu=b^\mu+p^\mu \tau$ and where we set retarded boundary conditions on the fluctuations, namely $z^\mu(-\infty)=0$. 
	In the KS gauge, the worldline action decouples trivially 
	\begin{equation}
		S[x;g] = S[x]_\eta +
		S[x]_h \, ,
		\label{split-action}
	\end{equation} 
	where the interactions  are now associated with  the second term in Eq.~\eqref{split-action}.
	The form of this action suggest that we can absorb the interacting part by a field redefinition.
	Let us consider again the expansion \eqref{expansion-action}.
It will be useful to consider its coefficients \eqref{coefficients} in energy and momentum space. Using our conventions 
	\eqref{convention-fourier}, we find that the first coefficient  in the expansion reads
	\begin{widetext}
		\be \label{linear-action}
		\frac{\delta S}{\delta z_\alpha(-\omega)}\Big|_{z=0} = -\frac{\ii}{2}\int \hat{\dd}^4\ell  \ e^{-\ii\ell\cdot b}\deltahat(\omega+\ell\cdot p) \hKS_{\mu\nu}(-\ell)\left( p^{\mu}p^{\nu}\ell^{\alpha}+2\omega p^{(\mu}\eta^{\nu) \alpha}\right) \, ,
		\ee
	\end{widetext}
	where we have used that $\delta S[x]_\eta/\delta z_\alpha=0$. This is nothing else but  the point particle current minimally coupled to background gravity
	\begin{equation}
		\begin{aligned}
			&J^\alpha_{1}(\ell,-\omega) \\
			&=
			-\frac{\ii}{2} e^{-\ii\ell\cdot b}\deltahat(\omega+\ell\cdot p) \hKS_{\mu\nu}(-\ell)\left( p^{\mu}p^{\nu}\ell^{\alpha}+2\omega p^{(\mu}\eta^{\nu) \alpha}\right) \, .
		\end{aligned}    
	\end{equation}
	The above term can be reabsorbed in the perturbative expansion of the configuration space variables in the action $S[x]$
	by performing a field redefinition in the path integral, namely
	\ba 
	\label{fr}
	\tilde z_{\mu}(-\omega) \to  & \tilde z_{\mu}(-\omega) + \tilde z'_{\mu}(-\omega) 
	\\ 
	= & \tilde z_{\mu}(-\omega) -\ii\int\dd\bar\omega \, \tilde G_{\mu\nu}(\omega,\bar \omega)J_{1}^\nu(\ell,\bar \omega) \, .
	\ea
	We stress that the above field redefinition does not generate any anomaly (functional determinant) since the path integral measure is translational invariant by definition. As we will see shortly we can use this field redefinition
	to perturbatively resum the path integral.
	The field redefinition \eqref{fr} allows us to get rid of the above linear action in the fluctuations, yielding to a redefined perturbative expansion for the particle deflection on the worldline, namely 
	\begin{equation}
		\begin{aligned}
			x^{\mu }(\tau) = & x_0^\mu + \int{\dd \omega} \ e^{\ii \omega \tau} {\tilde{z}}^{\prime\mu}(-\omega) + z^{\mu}(\tau) \\
			\equiv& x^{\mu, (1)}(\tau) + z^{\mu}(\tau)\, , 
		\end{aligned}
	\end{equation}
	which accounts for a $\mathcal O (G)$ (1PM) correction to the straight line of the point particle moving in the linearized Kerr background. 	
	From such correction now it is easy to derive the 1PM particle impulse in the background 
	\ba \label{lo-impulse}
	\Delta p^{\mu, (1)} =& \int\dd\tau \frac{d^2 x^{\mu, (1)}}{d\tau^{2}}  \\ = &
	-\frac{\ii}{2} \int \hat{\dd} ^4 q\  e^{-iq\cdot b} \deltahat(q\cdot p) \tilde{h}_{\alpha\beta}^{KS}(-q) p^{\alpha}p^{\beta}q^{\mu}\, ,
	\ea
	which exactly matches Refs.\cite{Vines:2017hyw,Arkani-Hamed:2019ymq, Menezes:2022tcs}. The expression of the momentum simplifies to 
	\begin{equation} 
		\begin{aligned}
		&	\Delta p^{\mu, (1)}  = 8GM m \pi \int \hat{\dd} ^4 q \  e^{-\ii q\cdot b} \deltahat(q\cdot p) \deltahat(q\cdot V)  \frac{q^\mu}{M X q^2} \\
&\quad \times (2 \gamma   \varepsilon (a,q ,p,V) \sin(X) +\ii \left(2 \gamma ^2-1\right) m M X \cos (X)) \, . 	
	\end{aligned}
	\end{equation}
	The integrals may be solved following the well-known strategy outlined in Ref.~\cite{Arkani-Hamed:2019ymq}. 
Specifically, using the identity
	\begin{equation}
		q^\mu \varepsilon (a,q ,p,V)= a\cdot q \  \varepsilon^{\mu} (q,p,V)\, , 
	\end{equation}
	and ignoring terms or order $q^2$ and higher in the integrand (these lead to local terms that vanish after integration)
	one finds 
	\newcommand{\bb}{b}
	\begin{equation}
		\begin{aligned}
	& \Delta p^{\mu, (1)}  =  \frac{2 G mM}{(\bb^2-|a|^2)\sqrt{\gamma^2-1}}
	 \\
	&\quad \quad\times\left((1-2\gamma^2)\bb+ 2|a|\gamma \sqrt{\gamma^2-1}      \right)\frac{-b^\mu}{\bb}.
			\end{aligned}
	\end{equation}

	The $1$\textrm{PM} all order in spin deflection angle then reads
	\ba\label{lo-angle}
	\alpha^{1\textrm{PM}} &= \frac{2 G M \left(2 |a| v
		-\left(v^2+1\right) \bb
		\right)}{v^2 \left(\bb^2-|a|^2
		\right)}
	\ea    
	in agreement with Ref.~\cite{Damgaard:2022jem}. Then, the small spin expansion can be obtained at any desired order
	\begin{equation}
		\begin{aligned}
			\alpha^{1\textrm{PM}} &= \frac{2 G M \left(v^2+1\right)}{v^2
				\bb }
			-|a|\frac{4 G M}{v  \bb ^2}
			+ |a|^2\frac{2 G M \left(v^2+1\right)}{v^2
				\bb^3}  \\
			&   -|a|^3\frac{4 G M}{v \bb ^4}
			+|a|^4\frac{2 G M \left(v^2+1\right)}{v^2
				\bb ^5} +\dots \,.
		\end{aligned}
	\end{equation}
	Note that under  the field redefinition the action $S_\eta[x]$ is modified as follows
	\begin{equation}
		\begin{aligned}
		\delta S_\eta[x]=& 
	-	\frac 12  \int \dd\tau \eta^{\mu\nu} \dot { z}'_\mu(\tau)  \dot { z}'_\nu (\tau)\\=&
	\frac{1}{2} \int \dd\tau \int \dd\tau'
	 z'_\mu(\tau) \eta^{\mu\nu}\frac{d^2}{d\tau^2}
	\delta(\tau-\tau') z'_\nu(\tau'),
		\end{aligned}
	\end{equation} 
which also affects the partition function \eqref{path-integral}  as
	\begin{equation}
		{\cal Z} \rightarrow  e^{\ii \chi_{\text{2PM}}} {\cal Z} .  
	\end{equation}
	Since the partition function is related to the eikonal phase through exponentiation, the above  must correspond to the $2$\textrm{PM} probe limit of the eikonal phase describing the scattering of a massive probe off linearized Kerr black hole, specifically: 
	\begin{widetext}
		\ba
		\ii \chi_{\text{2PM}} = -\frac{1}{2}\int d\omega_1 & \int d\omega_2\, J_{1\nu}(\ell,\omega_1)\tilde G^{\mu\nu}(\omega_1,\omega_2) J_{1\mu}(q, \omega_2) \\
		 = 		-\frac{1}{2}\int \hat {\dd}^4 q &\int \hat{\dd}^4 \ell e^{\ii b\cdot(q+\ell)} \hat{\delta} ((\ell+q)\cdot p) 
		\hat{\delta} (\ell \cdot p) \hat{\delta} (q \cdot p) \\
		& \times \Big(
		H(q)\cdot H(\ell) - \frac{H(q)\cdot\ell \, H(\ell)\cdot p}{2 p\cdot \ell} + \frac{H(\ell)\cdot q \, H(q)\cdot p}{2 p\cdot \ell}
		- \frac{H(\ell)\cdot p \, H(q)\cdot p \,\ell\cdot q}{4 (p\cdot \ell)^2}
		\Big)\,
		 ,\label{eikonal2PM}
		\ea
	\end{widetext}
	where we defined $H_\mu(q) = \tilde{h}^{KS}_{\mu\nu}(q)p^\nu$. This agrees with Ref.~\cite{Menezes:2022tcs}.
	
	Let us summarize what we have done. 
 We generated the integrands
	through  a field redefinition inside the functional integral \eqref{path-integral}.  This is equivalent to solve iteratively the point particle equation of motion but avoiding the evaluation of 
	the graviton corrections explicitly (see for instance  Appendix A of Ref.~\cite{Menezes:2022tcs}).
	This allowed us to get rid of terms linear in the quantum fluctuations from the interacting worldline action, thus directly generating the $1$\textrm{PM} correction to the particle deflection, which we used to evaluate the leading impulse \eqref{lo-impulse} for a probe scattering off linearized Kerr. But not only  the impulse, this technique allowed also to generate the integrand for the $2$\textrm{PM} probe limit of the eikonal phase related to such scattering event, by plugging in the functional integral \eqref{path-integral}, the field redefinition \eqref{fr}.
	The latter is nothing but the resummation of connected diagrams in the functional integral through the use of perturbative field redefinitions on the path integral variables.
	Such procedure can be iterated order by order in the \textrm{PM} expansion, and can be used to easily generate integrands for black hole scattering.
	We leave as a future investigation the study of the structure of higher order field redefinitions, which in principle could allow to get an exact resummation of the path integral. 
	\section{Back to the plasma}
	\label{nonhomogeneous}
	\subsection{The homogeneous case}
We define $\hat a= |a|/(GM)$ and	for the deflection angle we use 
	the notation $\alpha=\alpha_1+ \alpha_2+\dots $,  where $\alpha_i$ is proportional to $(G M/b)^i$. 
	The results of the previous sections are in agreement with
	Ref.~\cite{Crisnejo:2019ril}, where results up to $\mathcal{O}(G 
	^3M^3/b^3)$ are presented (see Appendix \ref{lowerorderangles}).  Notice that the spin is normalized with the
	gravitational radius $GM$ or the spin per unit mass in natural units as usual in GR, see e.g., Refs~\cite{Aazami:2011tw,Iyer:2009hq}.  This leads to an expansion that differs from the usual PM expansion .
	The linear in spin can be extracted from the probe limit of the angle computed in Ref.\cite{Jakobsen:2022fcj}. The higher orders in spin  can be obtained from Eq.~\eqref{angle-order} and the eikonal phase
	in Eq.~\eqref{eikonal2PM}.  The angle then reads
	\begin{widetext}
		\begin{align}
			\mathcal \alpha_4 = \left[
			\frac{105 \pi}{4} \left(
			\frac 1 {16} + \frac{1}{v^2}+\frac{1}{v^4}
			\right)- 
			12\left( \frac{1}{v^5}+\frac{10}{v^3}+\frac{5}{v} \right) \hat{a}
			+
			\frac{3\pi}{16}\left(15+\frac{72}{v^2} +\frac{8}{v^4}\right) 
			\hat{a}^2 
			-\frac{4 \hat{a}^3}{v}
			\right]
			\frac{G^4 M^4}{\bb^4} \, ,
		\end{align}
	\end{widetext}
	which can be compared against the results 	
	in Ref.~\cite{Jakobsen:2023pvx} taking the probe limit and directly to those in Ref.~\cite{Damgaard:2022jem}. In the massless limit ($v=1$) it matches Ref.\cite{Iyer:2009hq}.  
	Let us remark that although this angle can be recovered from the results in the literature, specifically from Ref.~\cite{Damgaard:2022jem}, here we have shown how it can be computed independently in worldline framework  bypassing the full computation. In particular, notice that all terms that appear in Eq.~\eqref{lo-angle} will contribute to the angle at all orders in the spin expansion. 
	
\begin{figure}[htb]
		\centering
		\includegraphics[width=0.48\textwidth]{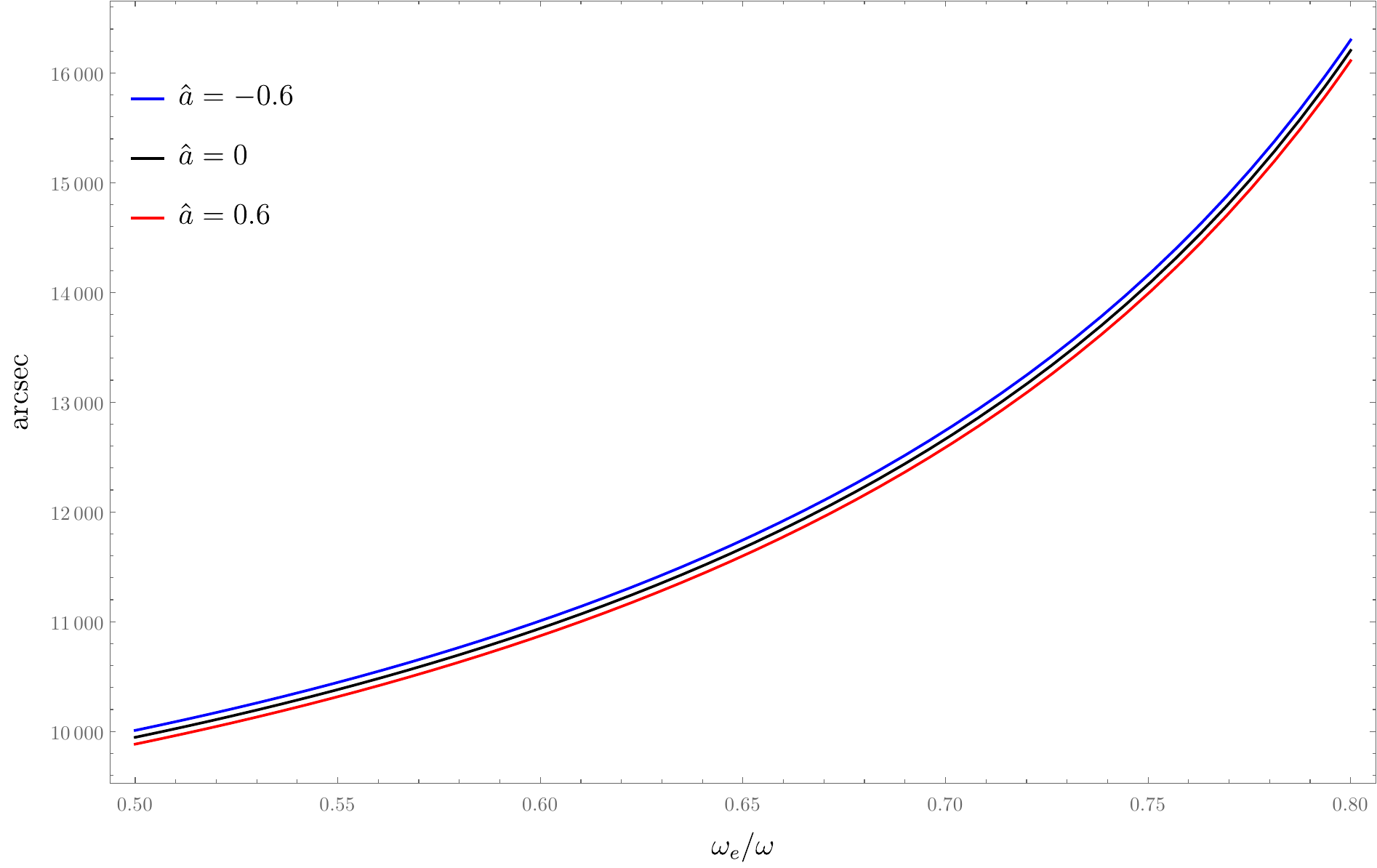}
		\caption{$\alpha$ as a function of $\omega_e/\omega$ in the homogeneous case.}
		\label{fig:homomegenous}
	\end{figure}

	Moreover, 	we can now compare the contributions in the case of a homogeneous plasma using the map \eqref{maptomedium}.
	 The contribution to the angle at this order is shown in Fig.\ref{fig:homomegenous}. We take $M$ to be the mass  of Sgr $A^*$, the black hole at the center of the galaxy $M=4\times 10^6 M_\odot$, while for the impact parameter we take $b=100 G M_\odot$. We plot three values of $\hat a$.
	\subsection{The inhomogeneous plasma}
	Let us now consider the case of an inhomogeneous medium in Schwarzschild.  In this case, the second term in the \eqref{action-position-space} also contributes. Specifically, upon the background expansion~\eqref{splitting-straight-line}, we obtain
	\begin{equation}
		\begin{aligned}
			S_{N}=- \ii \frac{k_e}{2\omega_0}
			\int \dd\tau \left[ N(x_0)+\partial_\mu N(x_0)z^\mu + \mathcal{O}(z^2)\right]
		\end{aligned}   
	\end{equation}
	at first order in $z^\mu$. We have used the short-hand notation 
	$k_e=4\pi e^2/(m_e)$. In the probe limit we can compute the impulse as 
	\begin{equation}
		\Delta p^{\mu}=-\omega_0 \int \dd \omega \,  \omega^2 \braket{\tilde z^\mu(\omega) \delta(\omega)} , 
	\end{equation}
	where we have used the equation of motion. At this order, the impulse separates into two contributions, which we label as
	\begin{equation}
		\Delta p^{\mu}=
		\Delta p_{\text{E}}^{\mu}
		+\Delta p_{\text{N}}^{\mu}\,,
	\end{equation}
	where $\text{E}$ labels the impulse in vacuum and $\text{N}$ on the medium. We have already computed the vacuum case so we focus on the latter. Upon expanding the action we have
	\begin{equation}
		\begin{aligned}
			&\Delta p_{\text{N}}^{\mu}\\
			=&	
			\frac{\ii k_e}{2}
			\int \dd \tau \int  
			\dd \omega \, \omega^2 
			\braket{\delta(\omega) \tilde z^\mu(\omega) ( N(x_0) + \partial_\nu N(x_0) z^\nu(\tau))  } \\
			=& \frac{\ii k_e}{2}
			\int \dd \tau  \partial_\nu N(x_0) \int  
			\dd \omega \, \omega^2 
			 \braket{\delta(\omega) \tilde z^\mu(\omega)   z^\nu (\tau)} \, .
		\end{aligned}
	\end{equation}
	In the last equality, we have used the fact that the worldline path integral of a single field vanishes. Recalling the expression of the propagator \eqref{propagator-energy-space} and expressing $z^\nu(\tau)$ in energy space, the last integration is trivial. 		We thus find 
	\begin{equation}
		\Delta p_{\text{N}}^{\mu}=
		\frac{k_e}{2\omega_0}
		\int \dd \tau  \, 
		\partial^\mu N(x_0)\, . 
	\end{equation}
	The evaluation of the impulse now requires the explicit form of   $N(x_0)$. Typical models are
	\begin{equation}
		\begin{aligned}
			N_E(r)=&N_0 \left(\frac  R {r}\right)^h, \\  N_S(r)=&N_0 e^{-r/R},	
		\end{aligned}\, 
	\end{equation} 
	which are appropriate for elliptical galaxy and spiral galaxies, respectively  \cite{Er:2013efa}. $R$ is the so-called scale radius.
 We will choose a frame where the particle moves along the $z$ direction with $r^2=\mathbf {b}^2+z^2$ and $b\cdot u=0$. In the approximation $\omega_{e0}/\omega_0 \ll 1$, we can also write $r^2\sim \mathbf {b}^2+\tau^2=\xo^2$.  Choosing the power law model we have  
	\begin{equation}
		\begin{aligned} 
			\Delta p_{\text{N}}^{\mu}= &
			\frac{N_0 k_e}{2\omega_0}
			\int_{-\infty}^\infty \dd \tau  \frac{N'_E(r)}{r}x_0^\mu \\
			= & b^\mu
			h\frac{N_0 k_e}{2\omega_0}
			\int_{-\infty}^\infty \dd \tau  \frac{1}{r^{2+h}}\,. 
		\end{aligned}
	\end{equation}	
Notice that the integral over the linear term in $\tau$ vanishes.		  We thus find
	\begin{equation}
		\begin{aligned} 
			\Delta p_{\text{N}}^{\mu}= &
			\frac{N_0 k_e}{\omega_0} \frac{\sqrt{\pi}
				\Gamma((h+1)/2)}{\Gamma(h/2)}\frac{b^\mu }{b} \left(\frac{R}{b}\right)^h \, .
		\end{aligned}
	\end{equation}
	The scattering angle can be computed from the impulse as
	\begin{equation}
		\alpha_N^{(0)}= \frac{|b\cdot \Delta p_{\text{N}}|}{b p_0}=   
		\frac{N_0 k_e}{\omega_0^2} \frac{\sqrt{\pi}
			\Gamma((h+1)/2)}{\Gamma(h/2)}\left(\frac{R}{b}\right)^h \, .
	\end{equation}
	At higher orders it is useful to introduce Feynman rules to represent the contribution at each order in $k_e$. The LO rule can be read off from the above computation. It reads
	\begin{equation}
		\raisebox{-0.5 em}{
		\begin{tikzpicture}[thick]
			\coordinate (A) at (0,0);
			\coordinate (B) at (1,-0);
			\node[very thick]{$\otimes$};
			\path[very thick,draw, dashed] (-1,0)--(-0.1,0);
			\path[very thick,draw] (0.1,0)node[above right] {$z^\mu(\tau)$}--(1,0) ;
		\end{tikzpicture}}
		=\ii h \frac{N_0 R^h k_e}{2\omega_0} \frac{x^\mu_0(\tau)}{r^{2+h}} \, .
	\end{equation}
	Similarly,  one can obtain from the action
	\begin{widetext}
		\begin{equation}
		\raisebox{-1 em}{
			    \begin{tikzpicture}[thick]
				\path [dashed, draw]
				(-1,0) -- (-0.1,0);
				\draw[very thick] (0,0.1) arc (-210:-280:1)node[right]
				{$z^\nu(\tau)$};
				\path [very thick,draw]
				(0.11,0) -- (1,0)node[right]
				{$z^\mu(\tau)$};
				\coordinate (A) at (0,0);
				\node[very thick]{$\otimes$};
			\end{tikzpicture}}
			=-\ii \frac{N_0 R^hk_e}{4 \omega_0 \Gamma(h) r^{h+4}}
			\left[(
			\Gamma (h+1)+\Gamma (h+2))x_0^{\mu }x_0^{\nu }-r^2 \Gamma (h+1) \eta^{\mu  \nu }
			\right] \,.
					\end{equation}
	\end{widetext}
	At NLO there
	is only one diagram with multiplicity 2 (originating from Wick contraction of four fields $z$) that contributes
	\begin{equation}
		F^\mu=\raisebox{-1.5 em}{
		\begin{tikzpicture}[thick]
			\coordinate (A) at (0,0);
			\coordinate (A1) at (0,-0.4);
			\coordinate (B) at (1,-0);
			\coordinate (B1) at (-1,-0.4);
			\node[very thick]{$\otimes$};
			\coordinate (C) at (0,0);
			\path[very thick,draw,dashed] (-1.5,0)--(-1.1,0);
			\path[very thick,draw] (-0.9,0)node[]at (-1,0) (c) {$\otimes$}--(-0.1,0);
			\path[very thick,draw] (0.1,0)node[above right]{$\omega=0$}--(1,0)node[right] {$\tilde z^\mu(\omega) \quad .$};
			\node (name) at (A1) {$t_2$}; 
			\node (name1) at (B1) {$t_1$}; 
		\end{tikzpicture}
	}
	\end{equation}
	Hence from the Feynman rules we find
	\begin{align}
		b_\mu F^\mu(t_1,t_2)=  &
		\frac{b^2 h^2 k_e^2 \Delta\left(t_1-t_2\right)  }{8 \omega _0^3 r\left(t_1\right)^{h+2} r\left(t_2\right)^{h+4 }}\nonumber \\
		& \times \left((h+2) \left(b^2+t_1 t_2\right)-r(t_2)^2\right),
	\end{align}
	where $\Delta(x)=|x|+x$.	Therefore, 
	\begin{equation}
		b_\mu \Delta p^\mu 
		=\int \dd t_1 \dd t_2  b_\mu F^\mu(t_1,t_2) \, .
		\label{bDeltap} 
	\end{equation}
We are interested in general values of $h$, which are not necessarily integer. For this purpose we use  analytic regularization and express the integrals as Gau{\ss } hypergeometric functions and later expand them around integer values of $h$.  The details are in  Appendix \ref{integrals}.
	The computation leads to
	\begin{equation}
		\alpha_N^{(1)}= \frac{|b\cdot \Delta p_{\text{N}}|}{\bb p_0}=  
		\frac{N_0^2 k_e^2}{2\omega_0^4} \frac{\pi^{3/2}
			\sec(h \pi)}{\Gamma(\frac 1 2-h)\Gamma(h-1)}\left(\frac{R}{b}\right)^{2h}\, ,
	\end{equation}
	which matches the results of Ref.~\cite{Crisnejo:2019ril} for the special cases  $h=1,2,3$. In addition, using the methods of Ref.\cite{Crisnejo:2019ril} we have used the Gau{\ss }-Bonet theorem to compute 
	 the angles for  $h=4,5,6$ finding agreement.
	 This representation is valid for all $h\ge1$ except for half integers $h$ where the secant has poles. The angle vanishes for $h=1$ but is nonvanishing for, say, $h=1.1$, which is relevant for phenomenological applications \cite{Psaltis:2011ru,Kimpson:2019mji}. 
	\section{Conclusions}
	\label{conclusions}
	In this paper we have studied light bending in the presence of the medium.  Using Synge's formalism, we constructed a position-space action and developed a corresponding worldline theory.
	
     The scattering angle can be computed from the probe limit using the well known analogy between the massive probe and  a  homogeneous plasma. Using the translation invariance of the worldline path integral, we have discussed how to extract the scattering angle using the saddle point approximation and field redefinitions.
     In the inhomogeneous case, where the analogy is not applicable anymore, we have considered a  power law model for the  plasma and derived a general expression for the NLO corrections due to the plasma.  It may be  interesting to explore the case of a dispersive medium \cite{Crisnejo:2019xtp, Bezdekova:2024vct} within the worldline formalism. 

	\section*{Acknowledgments}
	We thank Fiorenzo Bastianelli for discussions and Emanuel Gallo for helpful correspondence. 
	The work of LDLC  is  supported by 
	the European Research Council under grant ERC-AdG-885414. This research was supported by the Munich Institute for Astro-, Particle and BioPhysics (MIAPbP), which is funded by the Deutsche Forschungsgemeinschaft (DFG, German Research Foundation) under Germany's Excellence Strategy - EXC-2094 - 390783311.
	\appendix
	\section{Lower order angles}
	\label{lowerorderangles}
	We present the result of the deflection angles up to $\alpha_3$,  
	\begin{align}
		\alpha_1&= 2  \left(1 + \frac 1 {v^2}\right)  \frac{GM}{b}\, ,\\
           \alpha_2 &=  \left[3\pi  \left(\frac{1}{v^2}+\frac{1}{4}\right)-\frac{4 \hat a}{v}
           \right] \left(\frac{GM}{b}\right)^2\, ,\\
           \alpha_3 &= \Bigg[
          \frac{2}{3} \left(-\frac{1}{v^6}+\frac{15}{v^4}+\frac{45}{v^2}+5\right)
          -\frac{2 \pi  \hat  a \left(3 v^2+2\right)}{v^3} \nonumber
          \\ 
          &\qquad + \hat a^2 
           \left(\frac{2}{v^2}+2\right)
           \Bigg] \left(\frac{GM}{b}\right)^3 \, .
		\end{align}
	\section{Integrals} 
	\label{integrals}
The integration simplifies by expressing the propagator \eqref{time-domain-propagator} in terms of the Heaviside function 
\begin{equation}
	G^{\mu\nu}(t-t')=-\ii \frac{\eta^{\mu\nu}}{\omega_0}(t-t')\theta(t-t')\, ,
\end{equation}
where $\theta(x)=1$, for $x>1$, $\theta(x)=0$ for $x<0$ and $\theta(0)=1/2$. Introducing the change of variables $z_1=(t_1-t_2)$  write  Eq.~\eqref{bDeltap} as a linear combination of integrals of the form
 \begin{equation}
 	I[a,b]=\int_{-\infty}^{+\infty} \dd t \int_{0}^\infty \dd z_1 \frac{
 	z_1^a t^b }{ \left(b^2+t^2\right)^{\lambda_1} \left(b^2+(t+zz_1)^2\right)^{\lambda_2}
 }\,
 \end{equation}
where $a,b\ge 0 $  and  $\lambda_1=(h/2+1)-\epsilon$, $\lambda_2=(h/2+2)+\epsilon$. Here $\epsilon$ is an  analytic regulator. For positive integers $a,b$, the integral over $t$ can be evaluated introducing Schwinger parameters for each denominator and taking derivatives over $u$. For instance 
\begin{equation}
	\begin{aligned}
 I[1,2]= 
 	\int_{0}^\infty \dd^3 z
 \exp \left(-\left(b^2 \left(z_2+z_3\right)\right)-\frac{z_1^2 z_2 z_3}{z_2+z_3}\right) \\
 \times
 \frac{\sqrt{\pi } z_1 \left(2 z_1^2 z_3^2+z_3+z_2\right) z_2^{\lambda _1-1} z_3^{\lambda _2-1} }{2 \left(z_2+z_3\right){}^{5/2} \Gamma \left(\lambda _1\right) \Gamma \left(\lambda _2\right)} \, .
 \end{aligned}
 \end{equation}
 This integral evaluates to a linear combination of Gau{\ss } hypergeometric functions, which we can expand in powers of the regulator using HypExp \cite{Huber:2005yg}. The result reads
\begin{equation}
 I[1,2]=\frac{3 \pi ^{3/2} (h+1) b^{-2 h-1} \sec (\pi  h)}{h \Gamma \left(\frac{1}{2}-h\right) \Gamma (h+3)}	+\mathcal{O}(\epsilon)
\end{equation}	
The remaining integrals can be computed similarly. 
\bibliographystyle{JHEP}

\end{document}